\documentstyle[11pt,IAU207_pasp,twoside,psfig]{article}
\begin{document}
\title{Metal-rich and metal-poor globular clusters in ellipticals: Did
we learn anything?\\
{\it or} Constraints on galaxy formation and evolution from globular cluster
sub-populations}
 \author{Markus Kissler-Patig}
\affil{ESO, Karl-Schwarzschild-Str.~2, 85748 Garching, Germany}


\begin{abstract}
A brief review on globular cluster sub-populations in galaxies, and
their constraints on galaxy formation and evolution is given.
The metal-poor and metal-rich sub-populations are put in a historical context, 
and their properties, as known to date, are summarized. We review why the
study of these sub-populations is extremely useful for the study of
galaxy formation and evolution, but highlight a few caveats with the
current interpretations. We re-visit the current globular cluster system
formation scenarios and show how they boil down to a single scenario for
the metal-poor clusters (namely the formation in ``universal'', small
fragments at high $z$) and that a hierarchical formation seems favored for 
the metal-rich clusters. 
\end{abstract}


\section{History of blue and red populations}

\subsection{In the beginning...}
{\it Stellar sub-population} in galaxies were introduced with the concept of 
population {\tt I} and {\tt II} in the Milky Way (Baade 1944). 
Galactic globular clusters were typically associated with the population 
{\tt II}, although some were recognized as belonging to the disk/bulge 
(Becker 1950 and others). The idea of {\it globular cluster sub-populations} 
in the Galaxy was probably only brought to general acceptance with the work 
of Zinn (1985).

Ellipticals were not thought of as complex systems until the 80's and mostly 
viewed as hosting an old, metal-rich stellar population. The lack of
resolution into single stars still prohibits us from clearly identifying
stellar {\it sub-populations} in early-type galaxies. 
Only as spiral--spiral merger became an alternative to
forming ellipticals, Ashman \& Zepf (1992) came up with a simple model
that {\it predicted} globular cluster sub-populations in ellipticals if these
formed in major mergers. Shortly afterwards, Zepf \& Ashman (1993) 
identified such sub-populations in the color distributions of globular 
clusters in giant ellipticals. 

\subsection{Are mergers the answer?}

For several years, the idea of sub-populations associated with mergers raged 
within the community, still leading to hot debates today. The formation of
new globular(?) clusters was observed in merging galaxies (Holtzman et
al.~1992 for the first compelling evidence), further supporting the
merger scenario. The debate shifted to whether these new clusters would 
actually evolve into {\it globular} clusters (which now appears to be the 
case, e.g~Schweizer's contribution in this volume), and whether their general
properties (spatial distributions, total number, specific frequency, ages and
metallicities) are compatible with those of the metal-rich populations
in ellipticals (which is still an open question). Note the caveat that 
we implicitly assume that
what we learn from today's mergers applies to the mergers that potentially 
formed the giant ellipticals at much higher redshift, which might not be the 
case.

In the mid-90's, several groups started to propose alternatives to major mergers
being the only explanation for multiple sub-populations (e.g.~Kissler-Patig 
2000 and references therein). But the alternative ``scenarios'' remain fairly 
vague in terms of predictions or details.
Generally, all scenarios tend to mix to some degree globular cluster formation,
globular clusters system assembly, and galaxy formation. Section
\ref{models} attempts to provide a critical summary. 


\section{Sub-population properties: room to improve}

In order to understand the origin of the sub-populations (which is the
first step required in order to use them to constrain galaxy formation
and evolution), we need to understand their properties. Do the
sub-populations differ in metallicity only (metallicity differences
being the reason why we detected them in the first place)? Or do
they resemble each other in many aspects? A brief summary of our
current knowledge is given below. 

\subsection{Our biases}

In the mid-90's several {\it giant} ellipticals were known to host
sub-populations. Blue and red globular clusters were identified in the color 
distributions and sub-samples created based on the colors. A few
aspects/selections of these studies are of interest and lead to biases that
need to be understood as we try to understand the sub-population
properties in more detail. 

{\it i)} No ``clean'' sub-sample can be obtained from the 
color distributions alone; rather each sub-sample contains an unknown 
fraction of the other sub-population. That fraction is not trivial to
quantify since it depends both on observational errors that smear
each color peak, and on the necessary but too simple assumption that colors 
are driven by metallicity alone. {\it The fraction of contamination
should be estimated and the impact on the uncertainty of the derived properties
quantified.}

{\it ii)} The term {\it bi-modal} was introduced early on and is still
widely used, driving our thoughts automatically to {\it two}
sub-populations only. A third peak in the color distributions was
mentioned in some galaxies, but in order to study the properties of
individual sub-populations, only two groups were considered (true for
all studies to date). The latter assumption seem acceptable as a first
approximation since clear differences were noticed, but the question remains 
how long this perspective will limit our ability to understand systems
that are almost certainly more complex. {\it We should keep in mind that
not two but multiple sub-populations are most probably present, and
should work towards a finer splitting of the sub-populations}.

{\it iii)} Our samples of galaxies are still very much biased in favor
of central giant ellipticals, or at least very luminous giant
ellipticals. The reason is of course that these galaxies host the most
clusters and are therefore the easiest systems (with enough number
statistics) to be studied. Unfortunately, these are the galaxies that we
then tend to call ``typical'' despite the fact that they are the rarest
and most extreme examples. A much more subtle bias is introduced in
sub-population studies by an implicit selection of galaxies with well
separated peaks in their globular cluster color distribution. For these
cases the separation of the two sub-population is of course the easiest.
However, assuming that the metal-poor population is rather constant in
age and metallicity (see below) this selection does bias us against
galaxies with less metal-rich clusters and intermediate age metal-rich
sub-populations. {\it Future samples should include early-type galaxies of all
types and luminosities, and in all types of environments (especially 
intermediate-luminosity field galaxies), irrespective of the properties
of the globular cluster system}. 

\subsection{Differences and Similarities}

In the following we briefly list common properties and differences
between the metal-rich and metal-poor sub-populations, as studied to
date. No long description is given and the list of references is
restricted to a few studies that include further references (see also
Ashman \& Zepf 1998, Kissler-Patig 2000, Harris 2001 for further 
references).

\subsubsection{The spatial distributions} 

The first property that was studied separately for blue and red clusters
is the radial surface density profile. 
Geisler, Kim \& Lee (1996) first noticed, in NGC 4472, that
the red clusters were more concentrated towards the center than the blue
ones. The radial surface density profile of the red clusters is steeper
than the one of the blue ones leading to a color gradient with radius
for the whole system. This behavior was observed in the cluster systems
of several giant elliptical galaxies. 

Further, Kissler-Patig et al.~(1997) first noticed, in NGC 1380, that red and 
blue clusters also differed in their 2-dimensional spatial distributions. The
red cluster system appears more elliptical than the blue one. The blue
clusters are not only more diffuse but also follow a rather spherical
distribution, while the red clusters follow the ellipticity and position
angle of the galaxy. Again, this behavior was confirmed in a few other
galaxies.

This leads to an association of the blue clusters with the ``halo''
(although ill defined for early-type galaxies) and an association of the
red clusters with the ``bulge'' or spheroid, which represents the
majority of the stars in these galaxies (see also Sect.~4.).

\subsubsection{The kinematics} 

Kinematics for the red and blue sub-populations were investigated in
three early-type galaxies only (M~87, NGC 4472, NGC 1399), all central giant
ellipticals. It is thus unclear whether these can be regarded as
typical. Nevertheless, it is clear that the kinematics (both the
rotation and velocity dispersion) of the red and blue clusters differ in 
these galaxies. A consistent picture has not yet emerged (e.g.~Kissler-Patig 
\& Gebhardt 1998, Kissler-Patig et al.~1999, Zepf et al.~2000).
Notice also that no clear predictions exist for the kinematics of the
sub-populations as a function of assembly/formation scenario.

\subsubsection{Ages and metallicities}

The difference in metallicity between red and blue sub-populations is clear 
from the color distributions (metallicity being the main contributor to color 
at old ages). However, it remains unclear how much red and blue populations 
differ in age.  Relative age differences can be measured
spectroscopically (e.g.~Kissler-Patig et al.~1998, Cohen et al.~1998) or
photometrically (e.g.~Puzia et al.~1999). The bottom line is that red and
blue clusters appear coeval to within the (large) measurement errors
(2--4 Gyr), with some disputed claims of the red clusters being younger by a
couple of Gyr. The absolute ages of the clusters in early-type galaxies
appears similar to the one of the old clusters in the Milky Way and
M~31, as judged by spectroscopic line indices.

\subsubsection{The sizes}

Recently (Kundu \& Whitmore 1998, Puzia et al.~1999, Larsen et al.~2000)
size differences were discovered between red and blue clusters in
early-type galaxies. The blue clusters appear systematically more
extended in all galaxies, independently of radius. The currently favored
explanation is that the size difference is a relic of the formation
process: the blue clusters could have formed in a less dense environment
than the red ones.


\section{More than two sub-populations}

The above differences in sub-population properties confirm that a splitting 
in color leads to two sub-groups with physical differences. A question that was
seldom asked to date is: whether each sub-group really represents a
single sub-population, or whether one or both sub-groups actually host
multiple sub-populations.

\begin{figure}[th]
\psfig{figure=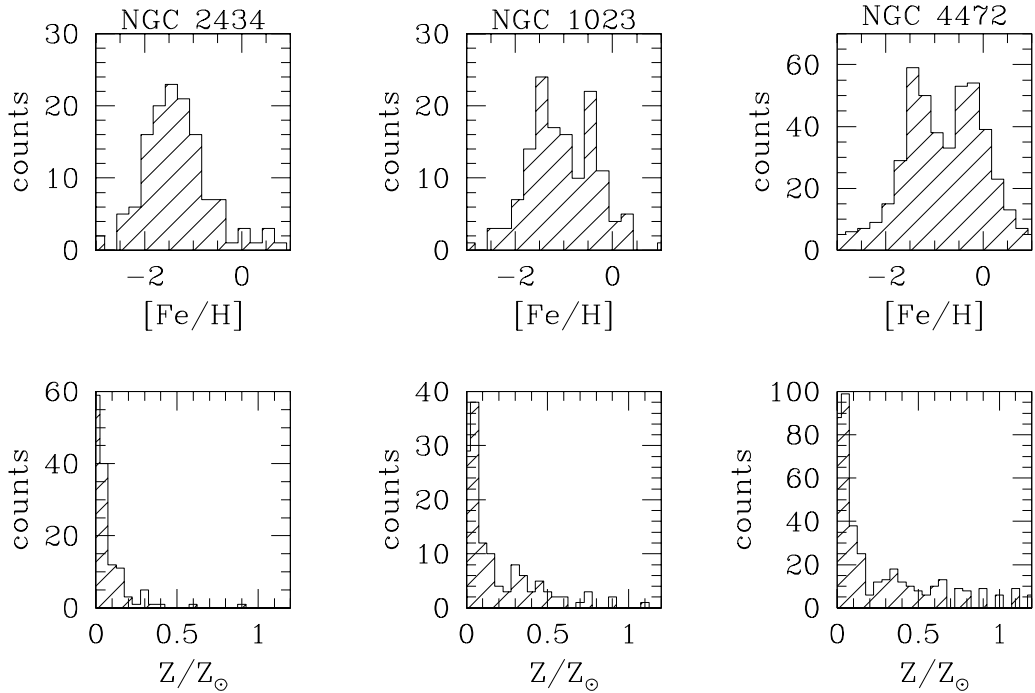,height=9.2cm,width=13.7cm
    ,bbllx=18mm,bblly=57mm,bburx=205mm,bbury=183mm}
\caption{Comparing [Fe/H] and Z distributions of globular clusters in
three early-type galaxies with a sequence of pronounced metal-rich
sub-populations.}
\end{figure}

The color (equivalent to [Fe/H]) histogram shows the distribution of a
logarithmic metallicity value. In figure 1, we illustrate for three cases
(with increasingly pronounced metal-rich population) how
such a distribution would look in linear metallicity Z. The two
clear peaks in [Fe/H] (translated linearly from $V-I$) disappear when
plotted in Z. For this particular choice of zero point (solar$=$1), the blue 
peak gets ``compressed'' into an even clearer peak between 0 and
0.1--0.2 Z$_\odot$.
The typical gap in color at [Fe/H]$\sim -0.7$ dex can still be guessed at
Z$\sim$0.2 . However, the red peak gets spread over several tenths in Z
and is not recognizable as a single peak anymore. Simple
simulations show that the spread of the metal-rich population is not
artificially induced by the fixed errors in logarithmic space, but
indeed due to a spread in metallicity that gets ``played down'' in [Fe/H].

While the blue clusters still appear to be a single physical group (see
also the next sections), there is no compelling evidence that this is
true for the red clusters too. Clearly, this prompts the question
whether the metal-rich sub-group is, in many or most cases, just an
amalgamation of multiple metal-rich sub-populations that do not
necessarily share the same origin.
Alternatively, for smaller red populations such as the one of the Milky
Way, the spread could also be explained by an extremely fast enrichment 
of the material out of which the metal-rich clusters formed (contrary to
the environment in which the metal-poor clusters formed). This would
imply that the metal-rich clusters show a small spread in age too, and
that they formed in a potential well deep enough to retain the enriched gas.
It remains to be shown whether the latter scenario could also apply to the
formation of several hundreds to thousands of globular clusters in giant
ellipticals.

Thus, the interpretation of the metal-rich peak appears more complicated
than originally assumed and the red population might well host multiple 
sub-populations.
This would imply a complex star/cluster formation history, i.e.~not a
single collapse or single major merger event might be at the origin of
the metal-rich clusters, but rather several such events. 
Given the above:
Does the result of a KMM test (that imposes 2 Gaussians to the color
distribution) make a physical sense? Or does it only mislead us to consider
two and not more sub-populations in a system, and thus over-simplify the
interpretation?


\section{The globular cluster -- star connection}

The above problems lead us to consider the connection between
globular clusters and stars. Can we identify the multiple stellar
populations that should be associated with the multiple globular cluster
sub-population?

\subsection{The link between globular cluster and star
formation}

A shaky aspect of our current interpretation of globular cluster
formation in terms of star-formation history of the galaxy is that we assume 
for simplicity that globular cluster formation traces star formation one-to-one.
The fact that a strong correlation exists between star and cluster formation is 
supported by the fairly constant (within a factor of two) specific 
frequency in all galaxies, the correlation between star and cluster formation, 
etc... (e.g.~Introduction in Puzia et al.~1999). However, a one-to-one
correlation is certainly too simplistic. In fact, there is growing
evidence that the blue globular clusters are associated with few stars,
while the red clusters form together with the majority of stars.

The first piece of evidence comes from the fact that high specific
frequencies are systematically observed in ``halos'' and in dwarf galaxies,
both dominated by blue clusters.

The second hint comes from the fact that diffuse stellar
population studies in early-type galaxies (Maraston \& Thomas 2000, Lotz
et al.~2000) imply only a small fraction (typically 10\%) of old
metal-poor stars as opposed to a majority of old metal-rich stars. In
contrast, old metal-poor and metal-rich clusters are roughly present
with a 50\%/50\% share in these galaxies, i.e.~blue clusters have fewer
stars associated with them than red clusters do.

The third indication comes from direct number counts of stars and
globular clusters as a function of metallicity in the nearby elliptical
NGC 5128 (see Harris, Harris \& Poole 1999). A comparison of the
metallicity distributions for stars and globular clusters shows the
much higher ratio of stars to globular clusters at the metal-rich end.

\subsection{The color of specific frequency}

The conditions for formation of the metal-poor globular clusters do not
seem favorable to the formation of a large fraction of stars, while the
contrary seems true for the metal-rich population. A possible
explanation could be that metal-poor clusters form in shallower
potential wells and eject a lot of gas during their formation that
cannot be processed further into stars. While the metal-rich clusters
form in deeper potential wells retaining the gas and allowing an efficient
star formation. This assumes that clusters collapse at the very
beginning of a star formation burst.

The fact is that the specific frequency ($Sn$) of the blue sub-population
is much higher than the one of the red sub-population. The specific
frequency of a galaxy could thus be increased by forming (or accreting)
a large quantity of metal-poor clusters$+$stars. On the other hand, when
comparing the specific frequency of different galaxies, one should
correct for the fraction of blue to red clusters present. Also the (in
my opinion wrong) argument that specific frequency has anything to do
with a major merger in the past history of the galaxy needs to be
revised. From the above, a major merger, producing metal-rich clusters,
could only decrease the specific frequency, and the specific frequency
of a spiral--spiral merger could be equal to or lower than for one of the 
progenitor spirals alone. But such arguments imply that we know the $Sn$
produced in mergers, which we do not.

In any case, the ``color of the specific frequency'' needs to be taken
into account when using $Sn$ to constrain any formation scenario, as
well as the fact that ``$Sn$'' is constant when normalized to total baryons
instead of just stellar light (McLaughlin 1999). 


\section{Formation models for dummies (by a dummy)}

Ideally, we would like to use globular clusters to constrain the star
formation history of their host galaxies. In practice, we need to know
how/where globular clusters form in the first place, and what the
assembly histories of the galaxies were. Information on all these points
is contained in the globular cluster system properties. However, all the
aspects are entangled and when we try to set up a ``scenario'' we often
mix all these aspects in.

The currently discussed scenarios fall roughly into three categories: {\it
i)} the formation of globular clusters and of the host galaxy in a major merger
event; {\it ii)} the formation of all globular clusters within the host
galaxy; {\it iii)} the formation of the globular clusters in fragments of 
different sizes that assemble later into a giant galaxy (e.g.~Kissler-Patig 
2000 for references to the different flavors of each scenario). Some
more scenarios combine all or some of the above.

We quickly summarize what we know about blue and red populations and
discuss these properties in the frame of the different scenarios.

\subsection{The Truth about blue and red populations}

\subsubsection{Are the blue and/or red clusters closely connected with the 
final host galaxy?}

\begin{figure}[th]
\psfig{figure=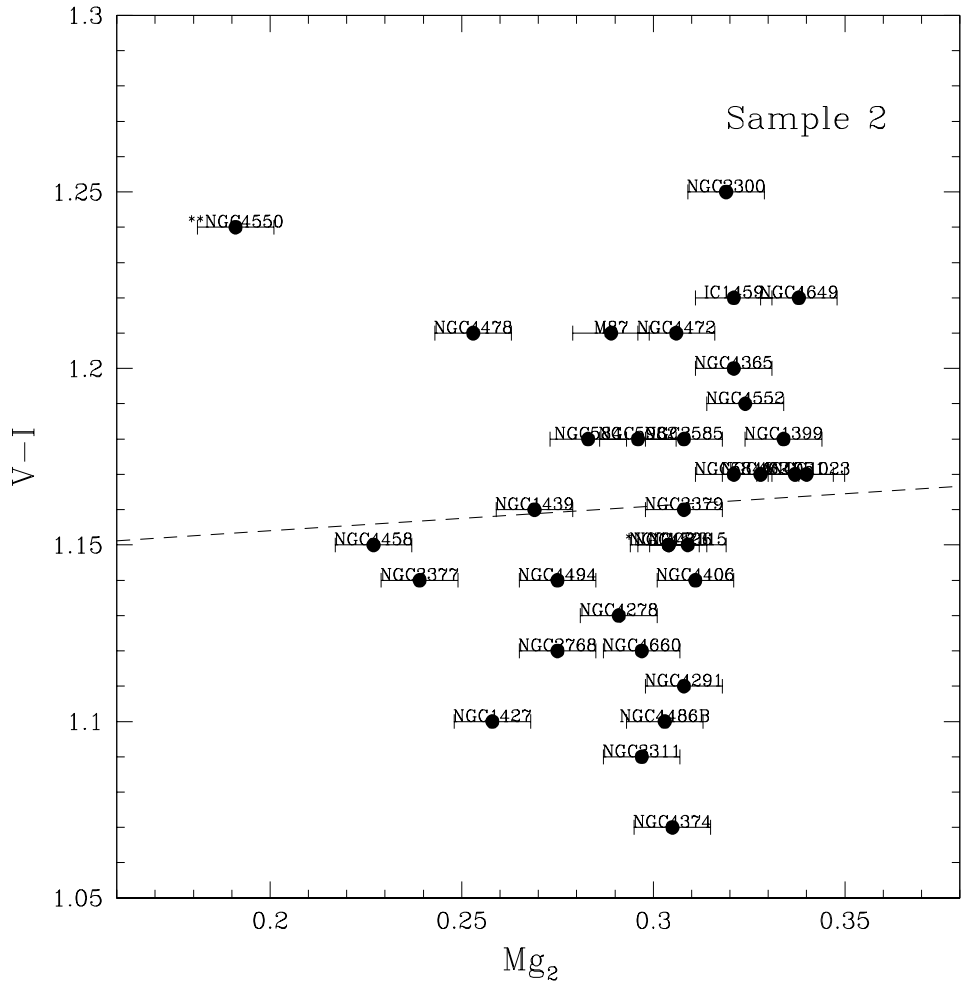,height=13cm,width=13cm
    ,bbllx=8mm,bblly=57mm,bburx=205mm,bbury=245mm}
\caption{Correlation between red peak color of the globular clusters and
Mg$_2$ index of the host galaxy, for a entire sample of clear and likely
``bi-modal'' systems, all values in the literature being used.
}
\label{correl}
\end{figure}

There are still too many Truths on blue and red populations. One recent
topic to review critically is the correlations of sub-population color
with host galaxy properties (Ashman \& Bird 1993, Forbes et al.~1997,
Burgarella et al.~2001, Forbes \& Forte 2001, Larsen et al.~2001). Only
a very weak correlation seems to exist between mean metallicity of the
blue sub-population and mean metallicity/size of the final host galaxy.
For the red sub-population a correlation appears more likely, although
different dataset give different results. As an example we plot in
Fig.~\ref{correl} (for all studies reporting a clear or likely ``bimodality'' 
in the globular cluster color distribution) the red peak color versus galaxy 
metallicity (Mg$_2$ index). All values in the literature were used, values 
for a same galaxy from different studies were averaged. The clear correlation
sometimes claimed for ``cleaned'' samples becomes less clear and more
complicated to interpret unambiguously (see Kissler-Patig et al.~2001
for more details and further samples). 

\subsubsection{The essence}

Despite all the uncertainties, we can retain several properties for each
sub-population.

For the {\bf blue} sub-populations, the following facts are reasonably
secure:\\
$\bullet$ The mean metallicity correlates only weakly if at all with the
host galaxy metallicity, and if present it is much shallower than
a one-to-one correspondence.\\
$\bullet$ Several properties (sizes, high $Sn$, ...) fit well with the idea of 
them having formed in shallow potential wells (small fragments).\\
$\bullet$ Their spatial distribution and kinematics favor them as being
``halo'' objects.

These clusters most probably formed in small ``universal'' fragments
(falling back on the Searle \& Zinn (1978) idea), some of them having formed 
with the dark matter potential of the galaxy (explaining a weak
correlation of the metallicities) but some having formed in satellites
and having been accreted later.

Concerning the {\bf red} population, we know that:\\
$\bullet$ The red color peak correlates more or less well with the
galaxy size and metallicity (although the correlation could still be
driven by biases in our samples)\\
$\bullet$ The red sub-population is likely to consist of several 
metal-rich sub-populations, driven by the probably  complex assembly history 
of the galaxy.\\
$\bullet$ These metal-rich clusters must have formed in denser
environments (mergers?) and deeper potential wells (larger fragments) than the 
metal-poor ones, given their sizes and the many stars associated with
them.\\
$\bullet$ The red globular clusters are as old (or almost) as the blue
ones.

These properties are still compatible with all red clusters having formed
in one collapse or one major merger event, but it appears more likely
that several such ``major'' events contributed to build up the giant
systems. The key to a discrimination will be the study of red
sub-populations in order to identify or rule out sub-populations within 
the metal-rich clusters.

\subsection{And the winner is ...}
\label{models}

Is one model currently favored over another? The main problem is that no
model makes a clear, unique prediction that would allow us to rule it out
or confirm it, and as stated earlier, these models often mix a number of
aspects.

Clearly, the current idea of the formation of metal-poor clusters is
compatible with all three flavors of scenarios. Thus, at early stages
(assuming that metal-poor material is related to an early epoch, which
is probably true in most cases) all scenarios burn down to one: namely small 
``universal'' fragments collapsing and assembling to ``halos'' with high
Sn (e.g.~see Burgarella et al.~2001 for a correspondence with the
high-$z$ universe). The main open question is: what fraction formed in
clear association with the final host galaxy and what fraction formed
independently and was accreted later?

Only for the metal-rich clusters, the scenarios might differ somewhat.
We need to remember that most clusters (and spheroids) assembled as
redshifts $z>1$ (probably $z>2$), at which stage the large collapse
leading to the formation of the bulge might well have been induced by a
gas-rich merger. The question thus is: did the red clusters form during
the single collapse of the spheroid or did they form in
multiple-collapses that assemble subsequently to form the
final host galaxy? Probably both, but which mode dominated? {\bf If} 
the red clusters really show distinct sub-populations, the latter, hierarchical
mode, would be favored. 


\acknowledgments
I wish to thank D.Burgarella, V.Buat, T.H.Puzia, P.Goudfrooij,
C.Maraston, D.Thomas for interesting discussions that inspired part of
the subjects discussed in this contributions.


\section*{Questions}

\noindent{\it Grillmair:\, } (on the color of $S_n$) If all stars formed
in clusters could differences between globular and stellar color (or [Fe/H])
distributions simply reflect different destruction rates? Old,
metal-poor clusters have higher angular momentum and are more protected
from bulge shocking etc, and contribute few field stars. Red clusters
orbit in more violent environments, are destroyed more often and
contribute lots of field stars. Comments? Do we need more spectroscopy
of field stars?

\noindent {\it Kissler-Patig:\, } I like your idea, although I cannot
comment on its likelihood. I guess that simulations of destruction rates
would be ``easy'' if we could feed them with realistic initial
conditions. Currently, not enough is known on red and blue cluster
kinematics to make a secure guess. Another way of testing this hypothesis
would be to compare in detail the abundances and abundance ratios in
stars and clusters. This will be a by-product of T.H.Puzia's thesis,
aiming at high S/N spectroscopy of clusters and stellar light in a
sample of galaxies. So we might know more in a couple of years.

\noindent{\it Elmegreen:\, } (on a similar topic) Is enough known about
the dispersed population of stars in clusters of galaxies (e.g.~color)
to identify them with one population or the other of globular clusters?

\noindent {\it Kissler-Patig:\, } I am afraid that our knowledge of the
dispersed stellar population in clusters is still extremely sparse. So
the answer is no. But as wide field cameras become available,
 the colors of the diffuse light is certainly an interesting way to
follow-up this question.

\noindent {\it Zepf:\, } The width of the peaks in [Fe/H] might be
dominated by observational uncertainties. In this case, when you take
the linear errors in [Fe/H] to an exponential to get Z, you naturally
produce a peak with a tail. So I think the inferred distribution in Z is
a little hard to interpret. A critical point in this regard that you
mentioned earlier are the kinematic difference seen between the red and
blue population.

\noindent {\it Kissler-Patig:\, } I agree that the photometric errors
are still significantly contributing to the width of the peaks and
produce part of the tail in the Z distribution. However, we conducted a
few simulations that showed that observational errors cannot explain all
the dispersion. Ideally, one wants to test this on a sample that is
not dominated by errors anymore. Such samples could either be very high
quality HST photometry, or infrared color distributions, were the
metallicity sensitivity is largely increased with respect to the
photometric errors. Such tests are currently being carried out by our group.

\noindent {\it Whitmore:\, } Zepf has already made most of the point I
wanted to make, but let me add that this also shows why we tend to get a
bi-modal distribution in most cases. The metallicity enhancement happens
relatively rapidly, so any distribution in time with an initial peak at
$\sim$ 15 Gyr will end up with a roughly bi-modal distribution in
[Fe/H]. However, I again caution about amplified noise. If you start
with 2 delta functions in age at say 15 and 10 Gyr, and then predict
[Fe/H], it would be similar to the observations. I agree with the
constancy of the blue peak and with a larger spread in the red peak,
which we also discussed in AJ, 114, 1797 (1997).

\noindent {\it Kissler-Patig:\, } I think that your argument boils down
to whether all globular clusters form faster than any enrichment process
(in which case you would not automatically expect a spread in Z and a
bimodal [Fe/H]) or whether the metal-rich clusters form over ``longer''
periods of time and can profit from the abundance enrichment of that
star-burst. But note that this latter case would be indistinguishable
observationaly from multiple metal-rich sub-population, unless these
exhibit a large spread in age (more than several Gyr).
An answer might come from the study of $\alpha$-elements in
metal-rich globular clusters, which probe the timescale of formation and
will reveal whether in a star-burst the clusters formed before or after the 
bulk of stars (another aspect of T.H.Puzia's thesis).

\noindent {\it Forbes:\, } CDM models fail in many ways to reproduce the
real universe. Any comment on how much we should trust them?

\noindent {\it Kissler-Patig:\, } Hierarchical clustering models might
fail to reproduce some features of the real universe, but they get quite
close in describing all essential aspects. Also, they are the best models we 
have so far. Of course, we can design formation scenarios for globular cluster 
system around hypothetic, sketchy models (nowhere described in details, not
making any concrete predictions), but would that make more sense? I think the 
important point for the above discussion is that {\it i)} for the blue
sub-population we do not need any galaxy formation model. We actually know 
enough on their properties and formation that we are in the ideal situation of 
being able to constrain any galaxy formation model, namely it should include
at an early stage ``universal'', metal-poor fragments of $10^6$ to
$10^{10}$ solar masses. {\it ii)} concerning the red sub-population, we
need to stay open minded, but from an observational side, large
structures appear to be rare at redshifts $z>1.5$, which roughly corresponds to
their redshift of formation. Thus, red clusters have likely formed in
larger structures than the metal-poor ones, but not in structures as
large as their final host galaxies.

\end{document}